\documentclass[acmsmall]{acmart}
\newcommand{\final}{1}

\usepackage{url}
\usepackage{booktabs}

\usepackage{pifont}

\usepackage{xcolor}
\definecolor{darkgreen}{rgb}  {0.0, 0.5, 0.0}
\definecolor{darkblue}{rgb}   {0.0, 0.0, 0.8}
\definecolor{lightblue}{rgb}  {0.0, 0.5, 0.6}

\usepackage{listings}
\definecolor{ListBGColor}{rgb}{0.9,0.9,0.9}
\definecolor{ListCommentColor}{rgb}{0,0.6.0}
\definecolor{ListKeywordColor}{rgb}{0,0,0.85}
\definecolor{ListPreprocessorColor}{rgb}{0.60,0.15,0.75}
\definecolor{ListStringColor}{rgb}{0.65,0.45,0.2}

\lstdefinestyle{customHLSL}{
  language=[ISO]C++,
  morekeywords={cbuffer, float, float3, float4, Vector, vec2, vec3, vec4, mat4, texture2D, sampler2D},
  alsoletter={\#},
  deletedelim=*[directive]\#,
  keywordstyle={[2]\color{ListPreprocessorColor}},
  morekeywords={[2]\#if, \#elif, \#else, \#endif, \#pragma, \#include, \#define},
}

\lstdefinestyle{customSelos}{
  language=[5.2]Lua,
  morekeywords={terra, var, quote, shader, vertex, fragment, code, input, varying, output, uniform, textureSampler, vif, features, virtual, override, feature, param, @UIType, sampler2D, vec2, vec3, vec4, mat4},
}

\newcommand{\intodo}    [1]{~\textbf{\color{red}TODO\@: #1}}
\newcommand{\newtxt}    [1]{#1}
\newcommand{\newtxttwo} [1]{{\color{cyan}#1}}

\newcommand{\revision} [1]{{\color{red}#1}}

\ifthenelse{\equal{\final}{1}}
{
  
  \renewcommand{\intodo}[1]{}
  \renewcommand{\newtxt}[1]{#1}
  \renewcommand{\newtxttwo}[1]{#1}
  \renewcommand{\revision}[1]{#1}
}
{}

\newcommand{\cppTitle}{\texorpdfstring{{C\nolinebreak[4]\hspace{-.05em}\raisebox{.17ex}{{++}}}}{C++}}

\newcommand{\cpp}{\texorpdfstring{{C\nolinebreak[4]\hspace{-.05em}\raisebox{.10ex}{{++}}}}{C++}}

\newcommand{\tildealias}{\raisebox{0.5ex}{\texttildelow}}

\newcommand{\secref} [1]{Section~\ref{sec:#1}}

\newcommand{\figref} [1]{Figure~\ref{fig:#1}}

\newcommand{\tabref} [1]{Table~\ref{tab:#1}}

\newcommand{\listref}[1]{Listing~\ref{list:#1}}

\acmSubmissionID{1017}

\setcctype[4.0]{by}

\setcopyright{cc}
\acmJournal{PACMCGIT}
\acmYear{2022}
\acmVolume{5}
\acmNumber{3}
\acmArticle{25}
\acmMonth{7}
\acmPrice{}
\acmDOI{10.1145/3543866}

\begin{document}

\title{Supporting Unified Shader Specialization by Co-opting \cppTitle{} Features}


\author{Kerry A. Seitz, Jr.}
\orcid{0000-0003-2424-9495} 
\affiliation{%
 \institution{University of California, Davis}
 \department{Department of Computer Science}
 \streetaddress{One Shields Avenue}
 \city{Davis}
 \state{CA}
 \postcode{95616}
 \country{USA}
}
\email{kaseitz@ucdavis.edu}

\author{Theresa Foley}
\orcid{0000-0002-2381-3544} 
\affiliation{%
 \institution{NVIDIA}
 \streetaddress{2788 San Tomas Expressway}
 \city{Santa Clara}
 \state{CA}
 \postcode{95051}
 \country{USA}
}
\email{tfoley@nvidia.com}

\author{Serban D. Porumbescu}
\orcid{0000-0003-1523-9199} 
\affiliation{%
 \institution{University of California, Davis}
 \department{Department of Electrical and Computer Engineering}
 \streetaddress{One Shields Avenue}
 \city{Davis}
 \state{CA}
 \postcode{95616}
 \country{USA}
}
\email{sdporumbescu@ucdavis.edu}

\author{John D. Owens}
\orcid{0000-0001-6582-8237} 
\affiliation{%
 \institution{University of California, Davis}
 \department{Department of Electrical and Computer Engineering}
 \streetaddress{One Shields Avenue}
 \city{Davis}
 \state{CA}
 \postcode{95616}
 \country{USA}
}
\email{jowens@ece.ucdavis.edu}

\renewcommand\shortauthors{Seitz, Foley, Porumbescu, and Owens}

\begin{abstract}
  Modern unified programming models (such as CUDA and SYCL) that combine host (CPU) code and GPU code into the same programming language, same file, and same lexical scope lack adequate support for GPU code specialization, which is a key optimization in real-time graphics.
Furthermore, current methods used to implement specialization do not translate to a unified environment.
In this paper, we create a unified shader programming environment in \cpp{} that provides first-class support for specialization by co-opting \cpp{}'s attribute and virtual function features and reimplementing them with alternate semantics to express the services required.
By co-opting existing features, we enable programmers to use familiar \cpp{} programming techniques to write host and GPU code together, while still achieving efficient generated \cpp{} and HLSL code via our source-to-source translator.

\end{abstract}

%
%

\ccsdesc[500]{Computing methodologies~Computer graphics}
\ccsdesc[300]{Software and its engineering~General programming languages}
\ccsdesc[100]{Software and its engineering~Compilers}
%
%

\keywords{Shaders, Shading Languages, Real-Time Rendering, Heterogeneous Programming, Unified Programming}

\maketitle

\section{Introduction}\label{sec:Introduction}

Real-time graphics programming is made more complicated by the use of distinct languages and programming environments for host (CPU) code and GPU code.
GPU code performs highly-parallel rendering calculations and is typically authored in a special-purpose shading language (e.g., HLSL~\cite{Microsoft:2014:SM5}, GLSL~\cite{Kessinich:2017:TOS}, or Metal Shading Language~\cite{Apple:2021:MSL}), while host code, which coordinates and invokes rendering work that uses this GPU code, is written in a general-purpose systems language (e.g., \cpp{}).
When using a shading language and its corresponding graphics API (e.g., Direct3D~\cite{Microsoft:2020:D}, Vulkan/OpenGL~\cite{Segal:2017:TOG, KhronosGroup:2016:V1A}, or Metal~\cite{Apple:2014:M}), programmers issue API calls to migrate data between host and GPU memory and to set up and invoke GPU code that uses this data.
They must ensure not only that data is transferred efficiently, but also that data availability and layout in GPU memory match what the GPU code expects.
Because host and GPU code exist in two separate programming environments, programmers are ultimately responsible for ensuring compatibility between host and GPU code, with little help from the graphics APIs.

In contrast, heterogeneous programming is simpler in a \emph{unified} environment, where both host and GPU code are written in the same language, can be in the same file, and share lexical scopes.
For example, in CUDA~\cite{NVIDIA:2007:CUDA}, developers write both host and GPU code in \cpp{}, and passing parameters and invoking GPU code looks essentially like a regular function call.
Similarly, programmers using SYCL~\cite{KhronosSYCLWorkingGroup:2021:S2S} author GPU code as \cpp{} lambda functions or as named function objects and invoke them using API functions, allowing both host and GPU code to coexist within a single \cpp{} function.
In these unified systems, host and GPU code can use the same types and functions and reference the same declarations.
Thus, these unified systems---by construction---avoid an entire class of compatibility issues that must be handled manually in graphics programming.\footnote{\newtxt{Similar to how compilers help keep .h and .cpp files in sync, a unified system extends this benefit to GPU code. Current shader programming models do not validate code across the host (.cpp) and GPU (.hlsl) boundary, whereas unified systems do.}} 
Because of the associated code reuse, compatibility, and ease-of-use benefits, our overarching goal is to enable development of unified shader programming systems that are practically useful for large-scale real-time graphics applications.

While CUDA and SYCL provide powerful unified programming models for General-Purpose GPU (GPGPU) computing, neither they nor other popular unified systems provide adequate support for GPU code \emph{specialization}.
Specialization is a pervasive and critically important optimization in real-time graphics---it can have a significant impact on runtime performance~\cite{Crawford:2019:SOI,He:2018:SLM,Seitz:2019:SMF}, major game engines create mechanisms specifically to support it~\cite{UnityTechnologies:2019:UUM,EpicGames:2019:UE4}, and game developers go to great lengths to enable it even in scenarios where it may not initially seem feasible~\cite{ElGarawany:2016:DLI}.
For this optimization, GPU code is compiled multiple times with different options to generate multiple compiled \emph{variants} of the original code, each specialized to a particular combination of features and configurations.
Then, at game runtime, host code selects which variants to invoke based on dynamic data such as information about the scene, the underlying hardware platform, and user settings.
The data necessary to decide which GPU variants to invoke is not available until runtime, but developers need to generate the specialized variants ahead of time because just-in-time compilation can increase game load times, can hurt performance during gameplay, and is disallowed on some platforms.

We make the key observation that expressing and implementing specialization requires coordination between \emph{specialization parameters} that are \emph{compile-time parameters} for GPU code but \emph{runtime parameters} for host code.
While this dichotomy can be worked around in systems that use separate environments and separate parameter declarations for host code and GPU code, it represents a fundamental tension in unified systems where the same parameter must serve both a compile-time and runtime role.

Because of its importance in real-time graphics, we aim to demonstrate how to provide support for specialization in a unified system.
Furthermore, to be practically useful today, such a unified specialization implementation must be feasible in an existing, widely used language in real-time graphics, and the effort required to build and maintain such a system must be sufficiently modest in order to be tractable for most design teams.
To this end, we present the following contributions:

\begin{itemize}
  \item The design of a unified programming environment in \cpp{} that provides first-class support for specialization by co-opting existing language features (\cpp{} attributes and virtual functions) and implementing them with alternate semantics; and
  \item A Clang-based tool\footnote{\url{https://github.com/owensgroup/UnifiedShaderSpecialization}} that translates code using our modified \cpp{} semantics to standard \cpp{} and HLSL code, compatible with Unreal Engine 4.
\end{itemize}

\newtxt{An explicit goal of this work is to show that a unified environment can be integrated into an existing, large-scale engine and coexist with its current shader programming system.}
\newtxt{As mentioned above, our implementation targets Unreal Engine 4 for this purpose.}
\newtxt{Also,} our current implementation focuses on compute shaders, but we believe the design of our specialization system is compatible with other shader types as well.
We discuss other shader types, along with other limitations and future work, in \secref{Limitations}.

\section{Related Work}\label{sec:RelatedWork}

\subsection{Implementing Specialization}\label{sec:ImplementingSpecialization}

Popular game engines like Unreal Engine 4 (UE4) and Unity implement GPU shader code specialization using preprocessor-based methods.\footnote{For additional background, please see the supplementary material.}
Programmers express specializations in GPU code using C-style preprocessor facilities (e.g., \texttt{\#ifs}), and then this GPU code is compiled multiple times with different \texttt{\#define} options to generate multiple \emph{shader variants} of the original code.
At runtime, host code selects which variant to invoke based on runtime variables that correspond to these \texttt{\#defines}.
Effectively, code in these systems has two representations of these \emph{specialization parameters}: a runtime variable for host code and a compile-time \texttt{\#define} for GPU code.

However, this technique is not viable in a unified environment, where we desire a unified representation for specialization parameters.
C-preprocessor directives are evaluated in the first step of compilation.
In non-unified systems, this compile-time-only technique can be used for GPU code because the host and GPU code exist in separate files and separate programming environments.
However, it does not work in host code because of the need to dynamically control shader variant selection based on runtime information.
Vulkan's ``specialization constants'' allow host code to modify the values of constants in GPU code at application runtime.
While this feature eliminates the need to specify all possible values for a specialization parameter upfront, it still uses distinct, non-unified definitions of these parameters in host code and GPU code.

\cpp{} templates, another common code specialization technique familiar to many programmers, are also inadequate for supporting GPU shader code specialization.
In \cpp{}, template parameters are evaluated at application compile time, resulting in multiple compiled versions of templated source code (analogous to compiling multiple shader variants).
However, templates are insufficient for expressing runtime decisions in host code, since all template parameter values must be available at application compile time.
In other words, templates provide compile-time polymorphism, which is desirable for GPU code, but host code needs runtime polymorphism instead.
\newtxttwo{This issue is illustrated by the CUDA sample discussed in \listref{CUDASampleOurs}, which shows how our design (introduced in \secref{DesignDecisions}) results in significantly shorter and more maintainable code compared to the template-based CUDA sample.}

\begin{lstlisting}[label=list:CUDASampleOurs,
  float=p,
  style=customHLSL,
  numbers=left,
  basicstyle=\ttfamily\bfseries\scriptsize,
  xleftmargin=0.14\textwidth,
  xrightmargin=0.14\textwidth,
  escapechar=|,
  morekeywords={Texture2D, SamplerState, RWTexture2D, uint, uint2, float2},
  literate={~} {$\sim$}{1},
  caption={
    Because our system provides first-class support for shader specialization, invoking GPU code from host code is significantly more maintainable compared to using \cpp{} templates for this task.
    The code in this Listing accomplishes the same task as lines 633--1025 of the reduction\_kernel.cu CUDA sample~\cite{NVIDIA:2022:CSR}.
    Invoking the templated GPU code in the CUDA sample requires \tildealias{}400 lines of code, with nested switch- and if-statements and significant repetition (variadic templates, mp\_list, and typelists do not fix this issue either).
    Our system (\secref{DesignDecisions}) can accomplish this same tasks with \tildealias{}30 lines of code (plus minor additional boilerplate), as shown in this Listing.
%
    \\\hspace{\textwidth}\hspace*{2ex}
%
    While \cpp{} templates support compile-time specialization of code, they are inadequate for implementing the shader specialization optimization because of the compile-time vs.\ runtime needs in GPU vs.\ host code, respectively.
    The problem stems from the need to map runtime values to compile-time template parameters, which results in a major maintenance burden.
    For example, adding support for 2048 threads in the CUDA sample requires adding a new \texttt{case} to each of the seven \texttt{switch (threads) \{...\}} blocks, which leads to further code duplication.
    In contrast, in our version shown above, adding a new option requires simply adding another integer to the specialization parameter (discussed in \secref{VirtualFunctions}) on line~\ref{line:CUDASampleOurs_numThreads}.
  }
]
// Boilerplate
template<class T>
class [[ShaderClass]] ReduceBase {
    [[specialization_SparseInt(1024, 512, ...)]] int numThreads;|\label{line:CUDASampleOurs_numThreads}|
    [[specialization_Bool]] bool nIsPow2;

    [[gpu]] virtual void reduceKernel(...) = 0;
};

// Original CUDA code, replaced by ShaderReduce6
template <class T, unsigned int blockSize, bool nIsPow2>
__global__ void reduce6(...) { ... }

// Our version (2 extra lines vs. reduce6)
template<class T>
class [[ShaderClass]] ShaderReduce6 : public ReduceBase {
    [[gpu]] virtual void reduceKernel(...) { ... }
};

// Boilerplate
template<class T>
class [[ShaderClass]] ReduceRunner {
    [[specialization_ShaderClass]] ReduceBase* reducer;

    [[entry_ComputeShader(reducer->numThreads, 1, 1)]]
    void runReducer(...) {
        reducer->reduceKernel();
    }
};

// The following replaces the ~400 lines of switch-/if-statements
template<class T>
void reduce(...) {
    ...
    ReduceBase* reduceShader;

    switch (whichKernel) {
        case 0: reduceShader = new ShaderReduce0<T>(); break;
        ...
        case 9: reduceShader = new ShaderReduce9<T>(); break;
    }

    reduceShader->numThreads = threads;
    reduceShader->nIsPow2 = isPowTwo(size);

    ReduceRunner runner;
    runner.reducer = reduceShader;
    runner.addComputePass(...);
}
\end{lstlisting}

In the Slang shading language~\cite{He:2018:SLM}, specialization options in GPU code are expressed primarily through interface-constrained generics, and host code interfaces with GPU code to generate specialized variants using a runtime reflection API\@.
Because host and GPU code exist in different environments, Slang sidesteps the compile-time vs.\ runtime dichotomy of specialization parameters.
Furthermore, popular host-code languages like \cpp{} do not support Slang-style generics/interfaces, which limits the ability to apply Slang's specialization methods in existing engines seeking to create unified systems.
\newtxt{Similarly, Rodent~\cite{Perard-Gayot:2019:RGR} utilizes partial evaluation~\cite{Futamura:1983:PCO} to generate specialized renderers for CPUs and GPUs, but widely used languages in computer graphics do not have such partial evaluation features.}

\subsection{Encapsulation of Shader Code}\label{sec:EncapsulationOfShaderCode}

\begin{sloppypar}
Several shading languages support encapsulation of shader code and parameters via object-orientation, including Cg interfaces~\cite{Pharr:2004:ISI}, HLSL classes~\cite{Microsoft:2018:IAC}, Spark~\cite{Foley:2011:SMC}, Slang~\cite{He:2018:SLM}, and the RenderMan Shading Language~\cite{Hanrahan:1990:ALF}.
Our design in \secref{ShaderClass} utilizes this approach and takes further inspiration from Kuck and Wesche~\shortcite{Kuck:2009:AFF}.
Their work implements an object model for GLSL that is managed by corresponding proxy objects in \cpp{}.
Whereas their system uses dynamic dispatch in GPU code (with optimizations to remove dispatch code when possible), ours guarantees static dispatch in generated GPU code.
More fundamentally, our work differs from these previous works by extending shader objects to include both GPU and host code, with unified representations of types, functions, and parameters.
\end{sloppypar}

Sh~\cite{McCool:2002:SM} implements shader programming as an embedded domain-specific language (DSL) in \cpp{}.
GPU shader code is expressed using special types and operators, meaning that host and GPU code use distinct syntax for things like control flow.
In contrast, our work uses regular \cpp{} for both host and GPU code, presenting a unified environment where host and GPU code use the same types and functions.
Additionally, Sh uses runtime metaprogramming to generate GPU code, whereas our system performs all code generation at compile time.

\subsection{Unified Shader Programming}\label{sec:UnifiedShaderProgramming}

BraidGL~\cite{Sampson:2017:SSF} and Selos~\cite{Seitz:2019:SMF} both present shader programming environments that meet our definition of ``unified,'' but neither BraidGL nor Lua-Terra~\cite{DeVito:2013:TAM} (the language in which Selos is written) are widely used languages in real-time graphics, such as \cpp{}.
\cpp{} does not support the \emph{static staging} feature that BraidGL uses to express shader code and specializations.
Similarly, Selos relies on \emph{staged metaprogramming}, defined as a key set of language features that are available in Lua-Terra, but \cpp{} lacks these key language features.
Rather than requiring that new features such as these be added to the underlying language, our approach focuses on co-opting existing language features to implement unified shader specialization.

New projects like Rust GPU~\cite{Embark:2021:RG} and Circle~\cite{Baxter:2021:CCS} allow programmers to author GPU shader code in general-purpose systems languages (Rust and \cpp{}, respectively).
Both of these projects aim to satisfy a necessary condition for unified shader programming---the ability to author both host and GPU code in the same language.
However, neither of these systems include language design provisions to allow dynamic logic in host code to influence compile-time specialization and selection of GPU code, which is central to supporting unified shader specialization.
While Circle is based on \cpp{}, it has departed significantly from the standard language by adding other language features, including new general-purpose metaprogramming features.
While it may be possible to build a cohesive specialization system using these new Circle features, our goal is instead to introduce as few syntactic and semantic changes to \cpp{} as possible, which both lowers development and maintenance costs and better enables programmers to write code that looks and feels like standard \cpp{}.

\section{Unified Specialization System Design}\label{sec:DesignDecisions}

The key insight of this work is that we can add support for specialization in a unified programming environment by co-opting existing features of a programming language and implementing them with alternate semantics to provide the services required.
Because our high-level motivation is to bring unified programming to existing, large-scale graphics applications, we must ensure that our methodologies can integrate with existing code and toolchains.
Thus, we have chosen to demonstrate our approach by building a unified system in \cpp{}, since it is the most widely used language in real-time graphics (as evidenced by its use in many game engines~\cite{EpicGames:2019:UE4, Linietsky:2021:GE, ElectronicArts:2021:FE, AmazonWebServices:2021:AL}).
In contrast, creating a new programming language or using an uncommon one would lead to increased development costs, because graphics developers would need to rewrite large portions of their codebases or write additional code to interface between languages.
Similarly, rather than adding new language features to \cpp{}, we have chosen to stay as close to standard \cpp{} as possible in order to avoid the potential maintenance costs of integrating arbitrary new features with future versions of \cpp{}.
Instead, our approach is to co-opt features already present in \cpp{}, thereby maintaining compatibility as the language continues to evolve.
While our current implementation utilizes \cpp{}, we believe that our key insight, as well as many of the ideas presented below, are transferable to other languages as well.

\begin{center}\ding{167}\end{center}

In the remainder of this section, we discuss the major design elements of our unified specialization system.
\listref{UnifiedShaderExample} shows an example shader written using our unified \cpp{}-based programming environment.
We explain the various parts of it in the next three sections.

\begin{lstlisting}[label=list:UnifiedShaderExample,
  float=t,
  style=customHLSL,
  numbers=left,
  basicstyle=\ttfamily\bfseries\scriptsize,
  xleftmargin=0.14\textwidth,
  xrightmargin=0.14\textwidth,
  escapechar=|,
  morekeywords={Texture2D, SamplerState, RWTexture2D, uint, uint2, float2},
  caption={An example shader using our unified shader system. A ShaderClass can contain both host and GPU code, written using standard \cpp{}11 syntax. Special \cpp{} attributes are used to express various shader-specific constructs (e.g., uniform parameters, specialization parameters, and entry point functions).}
]
class [[ShaderClass]] FilterShader {|\label{line:UnifiedShaderExample_ShaderClass}|
public:
  [[uniform]] Texture2D            ColorTexture;|\label{line:UnifiedShaderExample_UniformStart}|
  [[uniform]] SamplerState         ColorSampler;
  [[uniform]] RWTexture2D<float4>  Output;|\label{line:UnifiedShaderExample_UniformEnd}|

  [[specialization_ShaderClass]]|\label{line:UnifiedShaderExample_SpecializationStart}|
  FilterMethod* filterMethod;|\label{line:UnifiedShaderExample_FilterMethod}|

  [[specialization_SparseInt(2, 4, 8, 16)]]|\label{line:UnifiedShaderExample_SpecializationInt}|
  int IterationCount;|\label{line:UnifiedShaderExample_SpecializationEnd}|

  [[entry_ComputeShader(8, 8, 1)]]|\label{line:UnifiedShaderExample_numthreads}|
  void MainCS([[SV_DispatchThreadID]] uint2 DispatchThreadID) const|\label{line:UnifiedShaderExample_const}||\label{line:UnifiedShaderExample_Varying}|
  {
    float2 pixelPos = /* ... */;
    float4 outColor = ColorTexture.Sample(ColorSampler, pixelPos);

    for (int i = 0; i < IterationCount; ++i) {
      outColor *= filterMethod->doFiltering(pixelPos);|\label{line:UnifiedShaderExample_doFiltering}|
    }

    Output[DispatchThreadID] = outColor;|\label{line:UnifiedShaderExample_output}|
  }
};
\end{lstlisting}

\subsection{Use \cpp{} Attributes to Express Declarations Specific to Shader Programming}\label{sec:ShaderSpecificAttributes}

\begin{sloppypar}
In our system, programmers use \cpp{} attributes to annotate declarations related to shader-programming-specific constructs.
The attributes feature was introduced in \cpp{}11 to provide a standardized syntax for implementation-defined language extensions, rather than different compilers continuing to use custom syntaxes (e.g., GNU's \texttt{\_\_attribute\_\_((...))} or Microsoft's \texttt{\_\_declspec()}). 
Our implementation supports the following shader-specific attributes:
\end{sloppypar}

\begin{sloppypar}
\begin{itemize}
  \item \emph{Uniform parameters} are annotated using the \texttt{[[uniform]]} attribute (lines~\ref{line:UnifiedShaderExample_UniformStart}--\ref{line:UnifiedShaderExample_UniformEnd}).
  \item \emph{Specialization parameters} are indicated using the \texttt{[[specialization]]} set of attributes (lines~\ref{line:UnifiedShaderExample_SpecializationStart}--\ref{line:UnifiedShaderExample_SpecializationEnd}).
  We defer discussion of specialization to \secref{VirtualFunctions}.
  \item The \texttt{[[entry]]} set of attributes declares a function as the \emph{entry point} to use when invoking GPU code execution.
  For compute shaders, this attribute requires arguments for the thread group size (line~\ref{line:UnifiedShaderExample_numthreads}), similar to the \texttt{numthreads} attribute in HLSL\@.
  \item System-defined \emph{varying parameters} are attached to entry point function parameters using corresponding attributes, which are named following HLSL's convention (e.g., \texttt{[[SV\_DispatchThreadID]]} on line~\ref{line:UnifiedShaderExample_Varying}).
  \item Because our system unifies host and GPU code into the same file, all non-entry-point GPU functions must be annotated with the \texttt{[[gpu]]} attribute.\footnote{CUDA uses a similar approach, where GPU-only functions are annotated with \texttt{\_\_device\_\_} and functions that are callable from both host and GPU code with \texttt{\_\_host\_\_ \_\_device\_\_}.}  
  By manually annotating GPU functions, we can disallow or reinterpret certain language features in GPU code when appropriate, while continuing to allow host functions to freely use any language feature (see \secref{VirtualFunctions} for further discussion).
\end{itemize}
\end{sloppypar}

Using \cpp{} attributes to express elements specific to shader programming represents a departure from the intent of this language feature.
In general, non-standard attributes can be ignored by the compiler and, thus, should not change the semantics of a program.
However, our attributes are integral to correctly defining the semantics of shader code; ignoring these attributes will result in an incorrect program.
Nevertheless, attributes provide a clean and concise method for expressing the above concepts, so our system co-opts this language feature for unified shader programming.

\subsection{Modularize Host and GPU Shader Code Using Classes}\label{sec:ShaderClass}

To promote more maintainable coding practices, our design uses \cpp{} classes to modularize shader code.
Programmers declare that a class contains shader code using the \texttt{[[ShaderClass]]} attribute (line~\ref{line:UnifiedShaderExample_ShaderClass}).
As mentioned in \secref{EncapsulationOfShaderCode}, other systems have used object orientation to modularize shader code.
However, our ShaderClasses can contain both host and GPU code, which is a major departure from prior work.

Because of this unified design, host and GPU code reference the same shader parameter declaration.
Thus, these declarations are---by construction---always kept consistent in both host and GPU code, avoiding the need to maintain separate definitions.
Host code provides data to GPU code by assigning values to these parameters:

\begin{lstlisting}[style=customHLSL, numbers=none, xleftmargin=0.275\textwidth, xrightmargin=0.275\textwidth]
 FilterShader shader;

 shader.ColorTexture = colorTexture;
 shader.ColorSampler = colorSampler;
 shader.Output = outputTexture;
\end{lstlisting}

\noindent
Host code can also set shader parameters using methods defined within a ShaderClass (e.g., the class's constructor).

GPU methods within a ShaderClass must be declared \texttt{const} (line~\ref{line:UnifiedShaderExample_const}).
In general, GPU shader code cannot modify uniform and specialization parameters, so requiring that these methods be \texttt{const} imposes this restriction.
However, some uniform parameter types (e.g., \texttt{RWTexture2D}) allow modification from GPU code using specific operations, and our system does provide support for these operations accordingly (e.g., writing to the \texttt{Output} texture on line~\ref{line:UnifiedShaderExample_output}).

A ShaderClass may or may not be a complete, invocable shader program.
If a ShaderClass contains an entry point method, then it can be used as an invocable shader program.
However, programmers can also write a ShaderClass without an entry point method, allowing for encapsulation of functionality that can then be reused across different shader programs by using the ShaderClass as a member variable (line~\ref{line:UnifiedShaderExample_FilterMethod}).
Member variables of a ShaderClass type must be declared as specialization parameters, for reasons we discuss next.

\subsection{Implement Specialization by Co-opting Virtual Function Calls}\label{sec:VirtualFunctions}

\subsubsection{Basic Specialization Parameters}

Like uniform parameters, ShaderClasses also express specialization parameters as member variables that both host and GPU code can reference, providing explicit declarations of these parameters for both halves of shader code.
Therefore, our system can catch more errors at compile time than other systems where specialization parameters are implicit in GPU code.\footnote{E.g., UE4's system, as discussed in the supplementary material.} 

Host code can set these parameters based on runtime information using the same mechanisms that apply to uniform parameters, e.g.:

\begin{lstlisting}[style=customHLSL, numbers=none, xleftmargin=0.165\textwidth, xrightmargin=0.165\textwidth]
 FilterShader shader;
 shader.IterationCount = settings.getIterationCount();
\end{lstlisting}

While these parameters are runtime-assignable in host code, they must instead be compile-time-constant in GPU code to allow the underlying GPU code compiler to perform the optimizations that programmers expect when they use specialization.
Thus, to support specialization, the set of possible values for all specialization parameters must be statically available at compile time.
For some types (e.g., enums and bools), our system can determine these values automatically; for other types (e.g., ints), we follow UE4's approach by requiring that programmers manually enumerate the possible values (\listref{UnifiedShaderExample} line~\ref{line:UnifiedShaderExample_SpecializationInt}).
Using these options, our translator (\secref{Implementation}) can then statically generate all GPU shader variants of a ShaderClass at compile time, while still allowing host code to easily select which variant to invoke at runtime by assigning values to the specialization parameters based on runtime information.\footnote{To provide better error checking during development, our translator generates asserts to ensure that a specialization parameter's runtime value is one of the statically enumerated options. UE4 has similar error checking, but some other systems do not.}

This approach provides a simple mechanism that cleanly handles the runtime-for-host-code vs.\ compile-time-for-GPU-code requirements of specialization parameters.
However, by using class member variables for specialization parameters, it is not obvious how to conditionally declare uniforms and functions based on these parameters (e.g., the standard practice of using preprocessor \texttt{\#ifs} does not work in a unified system, as discussed in \secref{ImplementingSpecialization}). 
We solve this issue by allowing a ShaderClass to use another ShaderClass as a specialization parameter.

\begin{lstlisting}[label=list:FilterMethodImpls,
  float=t,
  style=customHLSL,
  numbers=left,
  basicstyle=\ttfamily\bfseries\scriptsize,
  xleftmargin=0.14\textwidth,
  xrightmargin=0.14\textwidth,
  escapechar=|,
  morekeywords={uint, uint2, float2, override},
  caption={ShaderClasses can contain virtual \texttt{[[gpu]]} methods. In GPU code, virtual function calls are converted from dynamic dispatch to static dispatch, generating multiple shader variants accordingly.}
]
class [[ShaderClass]] FilterMethod {
public:
  [[gpu]] virtual float4 doFiltering(float2 pos) const = 0;|\label{line:Filter_Base_doFiltering}|
};

class [[ShaderClass]] LowQualityFilter : public FilterMethod {
public:
  [[gpu]] virtual float4 doFiltering(float2 pos) const override|\label{line:Filter_Low_doFiltering}|
  {
    /* Low Quality Method */
  }
};

class [[ShaderClass]] MedQualityFilter : public FilterMethod {
public:
  [[gpu]] virtual float4 doFiltering(float2 pos) const override|\label{line:Filter_Medium_doFiltering}|
  {
    /* Medium Quality Method */
  }
};

class [[ShaderClass]] HighQualityFilter : public FilterMethod {
public:
  [[uniform]] int ExtraParameter; // specific to this class|\label{line:Filter_High_uniform}|
  [[gpu]] virtual float4 doFiltering(float2 pos) const override|\label{line:Filter_High_doFiltering}|
  {
    /* High Quality Method */
  }
};
\end{lstlisting}

\subsubsection{ShaderClass Specialization Parameters}

As shown in \listref{UnifiedShaderExample} on line~\ref{line:UnifiedShaderExample_doFiltering}, the \texttt{doFiltering()} function is provided by a member variable of type \texttt{FilterMethod} (\listref{UnifiedShaderExample} line~\ref{line:UnifiedShaderExample_FilterMethod}). 
\texttt{FilterMethod} is itself a ShaderClass, and it also has ShaderClass subtypes.
\listref{FilterMethodImpls} shows the implementations of these types.

The \texttt{doFiltering()} method is declared as a virtual method in the base \texttt{FilterMethod} class (\listref{FilterMethodImpls} line~\ref{line:Filter_Base_doFiltering}). 
Then, each subclass overrides this method to provide their own implementations (lines~\ref{line:Filter_Low_doFiltering},~\ref{line:Filter_Medium_doFiltering}, and~\ref{line:Filter_High_doFiltering}).
Based on runtime information, the host shader code can select which implementation to use in the \texttt{FilterShader}:

\begin{lstlisting}[style=customHLSL, numbers=none, xleftmargin=0.20\textwidth, xrightmargin=0.20\textwidth]
 FilterShader shader;
 QualityEnumType quality = settings.getQuality()
 if (quality == QualityEnumType::Low)
   shader.filterMethod = new LowQualityFilter();
 else if (quality == QualityEnumType::Medium)
   shader.filterMethod = new MedQualityFilter();
 else if (quality == QualityEnumType::High)
   shader.filterMethod = new HighQualityFilter();
\end{lstlisting}

In \cpp{}, virtual methods normally use \emph{dynamic dispatch}---at runtime, the method implementation that gets invoked depends on the runtime type of the variable.
However, in GPU shader code, \emph{static dispatch}---where the method that gets invoked is known statically at compile time---results in significant performance benefits.
This difference creates a conflict between host code and GPU code: host code needs to select which type to use based on runtime information, but GPU code should use static dispatch (which requires this type information at compile time) for optimal performance.

Therefore, when a ShaderClass uses another ShaderClass as a member variable, our system requires this variable to be a specialization parameter, which allows us to avoid dynamic dispatch in the generated GPU code.
Our translator generates different shader variants for each possible subclass of a ShaderClass-type specialization parameter in order to convert the virtual method calls into static function calls, thereby replacing dynamic dispatches with static dispatches.
At runtime, the correct shader variant is selected by using the runtime type of the specialization parameter.\footnote{Rather than using the built-in \cpp{} runtime type information feature, we use our own, simplified mechanism to minimize performance overheads.}
By co-opting virtual functions and implementing them with alternate semantics for shader code, we are able to provide first-class support for GPU code specialization in our unified shader programming environment.

\begin{sloppypar}
As an added benefit, this design also encourages more robust software engineering practices.
For example, in \listref{FilterMethodImpls}, the \texttt{ExtraParameter} uniform parameter (line~\ref{line:Filter_High_uniform}) only applies when using the high-quality filter method.
Because this parameter is encapsulated within the \texttt{HighQualityFilter} class, it cannot be accessed elsewhere by mistake.
In contrast, standard practice in other systems would be to declare this parameter under a preprocessor \texttt{\#if}.
If other parts of the HLSL code need to access this parameter, programmers can (and often do) write additional \texttt{\#if} checks before using the parameter.
This practice leads to difficult-to-maintain code, since these various dependencies can be scattered throughout a large HLSL file.
Our design not only brings the specialization optimization to a unified environment but also enables shader programmers to utilize more features of \cpp{} to organize their code, rather than relying solely on the limited feature sets provided in standard shading languages.
\end{sloppypar}

While our current implementation always converts virtual functions to use static dispatch in GPU code, our design intentionally leaves open the possibility of compiling this same code to use dynamic dispatch instead.
For example, our implementation could generate conditional statements to dynamically select which function to invoke, or if HLSL adds support for virtual functions in the future, our system could compile directly to that feature.
Dynamic dispatch can reduce the number of compiled shader variants and is also important for real-time ray tracing.
In addition, generating partially specialized variants that utilize both static and dynamic dispatch can improve performance in deferred rendering~\cite{ElGarawany:2016:DLI,Seitz:2019:SMF}.
Alternatively, relying on \cpp{} template metaprogramming for shader specialization would necessarily always generate fully specialized, static shader variants, thereby limiting future adaptability of the system \newtxttwo{(in addition to the issues discussed in \listref{CUDASampleOurs})}.
Thus, co-opting virtual functions for specialization provides the flexibility to support future scenarios that will be important in real-time rendering.

\section{Translation Tool Implementation}\label{sec:Implementation}

To implement our design, we built a source-to-source translator based on Clang.
The translator uses Clang's LibTooling API~\cite{TheClangTeam:2022:L}, which provides a high degree of flexibility and power without requiring modifications to Clang.
Because our implementation is external from the Clang codebase, we can more easily update to newer Clang versions in the future to remain compatible with future \cpp{} features.
In addition, we use {HLSL\nolinebreak[4]\hspace{-.05em}\raisebox{.10ex}{{++}}}~\cite{Lopez:2022:h} to provide definitions of HLSL-specific types and intrinsics in \cpp{}.

The main task of the translator tool is to convert unified \cpp{} shader code that uses our co-opted features into standard \cpp{} and HLSL code that implements the alternate semantics for these features.
This transformation lets our system use existing \cpp{} and HLSL compilers and toolchains for final executable code generations, rather than requiring a full compiler implementation.
By using this translation strategy, we better facilitate ease of integration into existing applications.
Our translator is separated into three major components: the frontend, the host backend, and the GPU backend.

\subsection{Frontend}

The translator's frontend traverses the Clang Abstract Syntax Tree (AST) to retrieve relevant information from user-written source code.
Rather than operating on arbitrary regions of the AST, the frontend only inspects \cpp{} declarations that are annotated with the \texttt{[[ShaderClass]]} or \texttt{[[gpu]]} attributes.
An internal representation is created for each ShaderClass that contains information about its shader-specific elements (\secref{ShaderSpecificAttributes}).
Our translator operates on each \cpp{} translation unit individually, creating internal representations for all ShaderClasses and GPU functions within.
Then, our host and GPU backends use these internal representations to generate UE4-compatible \cpp{} and HLSL code, respectively.

\subsection{Host Backend}

\newtxt{Our current implementation outputs code that utilizes UE4's macro and render graph systems under the hood, in order to easily integrate with existing UE4 code.}
\newtxt{Other game engines and renderers could also be supported by writing additional backends for them using a similar approach.}

The host backend generates one or more UE4 Global Shader class implementations (hereafter referred to as an \emph{ImplClass}) for each ShaderClass.
These generated ImplClasses use UE4's macro system to implement the host-side representation of a ShaderClass's uniform parameters, as well as its boolean-, integer-, and enum-type specialization parameters.
If a ShaderClass has no ShaderClass-type specialization parameters, then only one ImplClass is generated.
To support ShaderClass-type specialization parameters, the translator generates multiple ImplClasses based on all possible combinations of runtime types for each such parameter.
For example, the shader in \listref{UnifiedShaderExample} would result in three ImplClasses, one for each \texttt{FilterMethod} subtype.

In addition, the translator generates code to interface user-written ShaderClasses with their underlying ImplClass implementations.
This task includes selecting which ImplClass to use based on the runtime types for each ShaderClass-type specialization parameter (if applicable), as well as communicating uniform and basic-type specialization parameters to their underlying UE4-based implementations.
Thus, while our system uses UE4's under the hood, programmers do not need to interact with this underlying implementation directly.
Instead, they can simply use the features provided by our unified system.

\subsection{GPU Backend}\label{sec:GPUBackend}

\newtxt{Similar to the host backend, our implementation currently targets HLSL for GPU code, but it could support other shading languages like GLSL via additional backends or by cross-compiling HLSL to another language (UE4 uses the latter approach).}
Our translator's GPU backend outputs an HLSL file for each ShaderClass with an entry point function.\footnote{ShaderClasses without entry point functions are not invocable shader programs, so outputting HLSL files for them is unnecessary.}
A ShaderClass's generated HLSL file contains all of the GPU shader code needed for every ImplClass of that ShaderClass.
This includes all uniform parameters and GPU functions from both the main ShaderClass and all ShaderClasses that it uses as specialization parameters (and their subtypes).
Any code that is specific to an ImplClass (e.g., the code specifically for each \texttt{FilterMethod} mentioned above) is output under a distinct \texttt{\#if} for that ImplClass.
When generating executable kernel code from these HLSL files, each ImplClass supplies the proper \texttt{\#define} option to the underlying HLSL compiler, ensuring that the generated shader variant is specialized to only the code it needs.

Our implementation also supports writing hardcoded HLSL directly within ShaderClasses and GPU functions.
This code is copied to the output HLSL files verbatim.
This feature serves two practical purposes.
Primarily, it lowers the barrier to porting shader code to use this system by allowing programmers to rewrite existing HLSL code incrementally, which better enables existing systems to adopt a unified shader design.
Secondarily, while our backend does convert some \cpp{} code to HLSL, not all HLSL features are supported, nor do all \cpp{} features translate to HLSL code properly.
By supporting hardcoded HLSL in our current implementation, we are able to explore unified shader specialization without first implementing every HLSL feature in \cpp{}, and vice versa, as a prerequisite.
\newtxt{Furthermore, the task of \cpp{}-to-GPU-shader-code translation is already being explored by Circle~\cite{Baxter:2021:CCS}.}

\section{Evaluation}\label{sec:Evaluation}

To evaluate our design, we ported shaders from UE4 to our system.
\newtxt{These shaders are complex, multi-platform, highly optimized, production shaders,}
\revision{and our ported versions are fully featured, drop-in replacements for their UE4 counterparts.}
Because feature-complete \cpp{}-to-HLSL translation is out of scope for this work, we use hardcoded HLSL code (\secref{GPUBackend}) in some parts of our ported code.
All results were obtained using UE4 version 4.25.4 built from source.\footnote{We used the release branch at commit b1e746725e8e540afe7ac586496b4ee4c081a10e.}
Since the unified shaders contain both host and GPU code, we rebuilt the modified files accordingly prior to benchmarking the ported code.
We review our findings below.

\subsection{ShaderClass Modularity}

Qualitatively, we have observed that using ShaderClasses to modularize shader code and specialization options leads to code that is more maintainable and easier to understand.
For example, UE4's temporal anti-aliasing shader\footnote{\newtxt{Interested readers can register for free access to UE4 source code and view this shader in the UE4 repository at \url{Engine/Shaders/Private/PostProcessTemporalAA.usf}.}} has many different specialization parameters usages scattered throughout the GPU code.
One such parameter chooses between different methods to use for caching texture reads.
These caching methods all rely on the same set of uniform parameters, and these particular uniforms are not used elsewhere in this shader.
However, all uniform parameters are declared as global variables, so this structure is not evident from looking at the code.
In our unified version of this code, we create a base ShaderClass to encapsulate these caching-specific uniform variables, and then each caching method inherits from this base class.
Thus, the code dependencies are readily apparent.

Furthermore, code reuse is made simpler and more apparent in our version of this shader as well.
Our base ShaderClass declares four virtual functions and provides implementations for them.
One subtype overrides all four functions, while another only overrides two of them (and reuses the base class implementations for the other two).
Again, this code structure is plainly evident in our version of this shader, but uncovering this underlying structure from the original UE4 source code required spending significant time tracking dependencies across \tildealias{}500 lines of HLSL code (within a \tildealias{}2,000 line file).
\newtxt{This shader (and many others) is riddled with \#ifs throughout, making it very difficult to understand and modify.}
\newtxt{This practice is standard in the industry.}
\newtxt{In contrast, our system enables object-oriented organization of shader code---while still supporting shader specialization---reducing the need for the \tildealias{}225 \#if/\#elif/\#else lines in the original shader.}
\newtxt{We believe organizing code using modern programming language features is vastly preferable to hundreds of scattered \#ifs.}
Our design enables users to cleanly modularize their shader code using \cpp{} classes and virtual functions, while still resulting in specialized shader variants that graphics programmers expect.

\subsection{Lines of Code}

Since our system design utilizes various abstractions for shader programming, we want to verify that these abstractions do not lead to excess code bloat.
\tabref{LOC_UE4vsUnified} compares the lines of code (LOC) for our rewritten shaders against the corresponding original UE4 code.
In UE4, an HLSL file can contain code for multiple shader programs; however, we have not necessarily ported all shader programs within an HLSL file to use our system.
To present a fair comparison, we only count lines of HLSL code related to the shader programs we have ported.

As shown, the LOC counts for the unified shader code are comparable to the original code.
Some of the additional lines in the unified code come from stylistic choices (e.g., putting the \texttt{[[gpu]]} function attribute on its own line).
Furthermore, some lines come from temporary code duplication.
Because we have not ported all UE4 HLSL files to our system, some code in our unified files is duplicated from HLSL header files that were \texttt{\#included} in the original shader code.
\revision{We manually copied and ported the necessary code segments from the header files into our unified shader files, so these code segments are counted in the LOC numbers for the unified versions of the shaders.}
\revision{We copied 7 LOC for the Motion Blur Filter shader and 5 LOC for the Temporal AA shader.}
\revision{The original UE4 shaders \texttt{\#included} this code, so we count the \texttt{\#include} lines for UE4 but not the code segments in the headers.}
While this duplication is ideally temporary, programmers still need to manage this code as a necessary overhead when incrementally porting large systems.
We believe the benefits of a unified system outweigh this extra temporary overhead, especially given that unified programming can reduce code duplication by allowing host and GPU code to share types, functions, and parameters.

\begin{table}[t]
\centering
\caption[Lines of Code]{Lines of code (LOC) for original UE4 shader code vs.\ the versions ported to our unified system.
We report only non-commented, non-empty lines, as reported by cloc~\cite{Danial:2022:c}.
The UE4 LOC number for each shader includes both the \cpp{} file (host code) and the corresponding HLSL file (GPU code), while the unified code uses a single file for both host and GPU code.
}\label{tab:LOC_UE4vsUnified}
\begin{tabular}{lcc}
\toprule
\textbf{Shader}       & \textbf{Original UE4 Code}  & \textbf{Our Unified Code}\\ 
(lines of code)       & \cpp{} file \& HLSL file    & \cpp{} file\footnotemark{}\\
\midrule
Motion Blur Filter    & 902                         & 920\\
Temporal AA           & 2,138                       & 2,251\\
\bottomrule
\end{tabular}
\end{table}
\footnotetext{The unified \cpp{} file includes some hardcoded HLSL code, since full \cpp{}-to-HLSL translation is out of scope for this work. This embedded HLSL code is included in the LOC counts.}

\subsection{Performance}

\begin{figure}[b]
\centering
  \includegraphics[width=0.48\columnwidth, height=0.27\columnwidth]{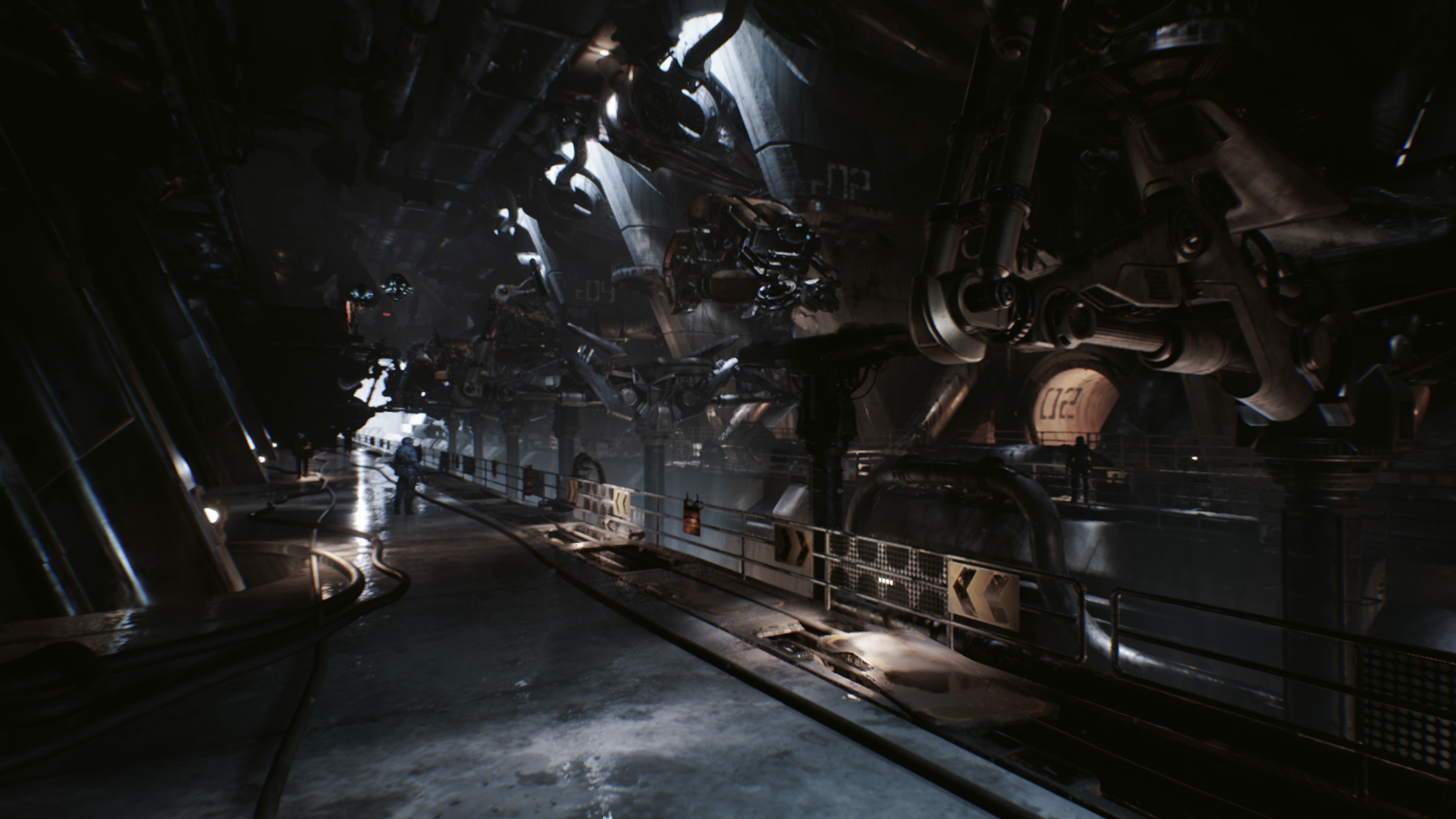}
  \includegraphics[width=0.48\columnwidth, height=0.27\columnwidth]{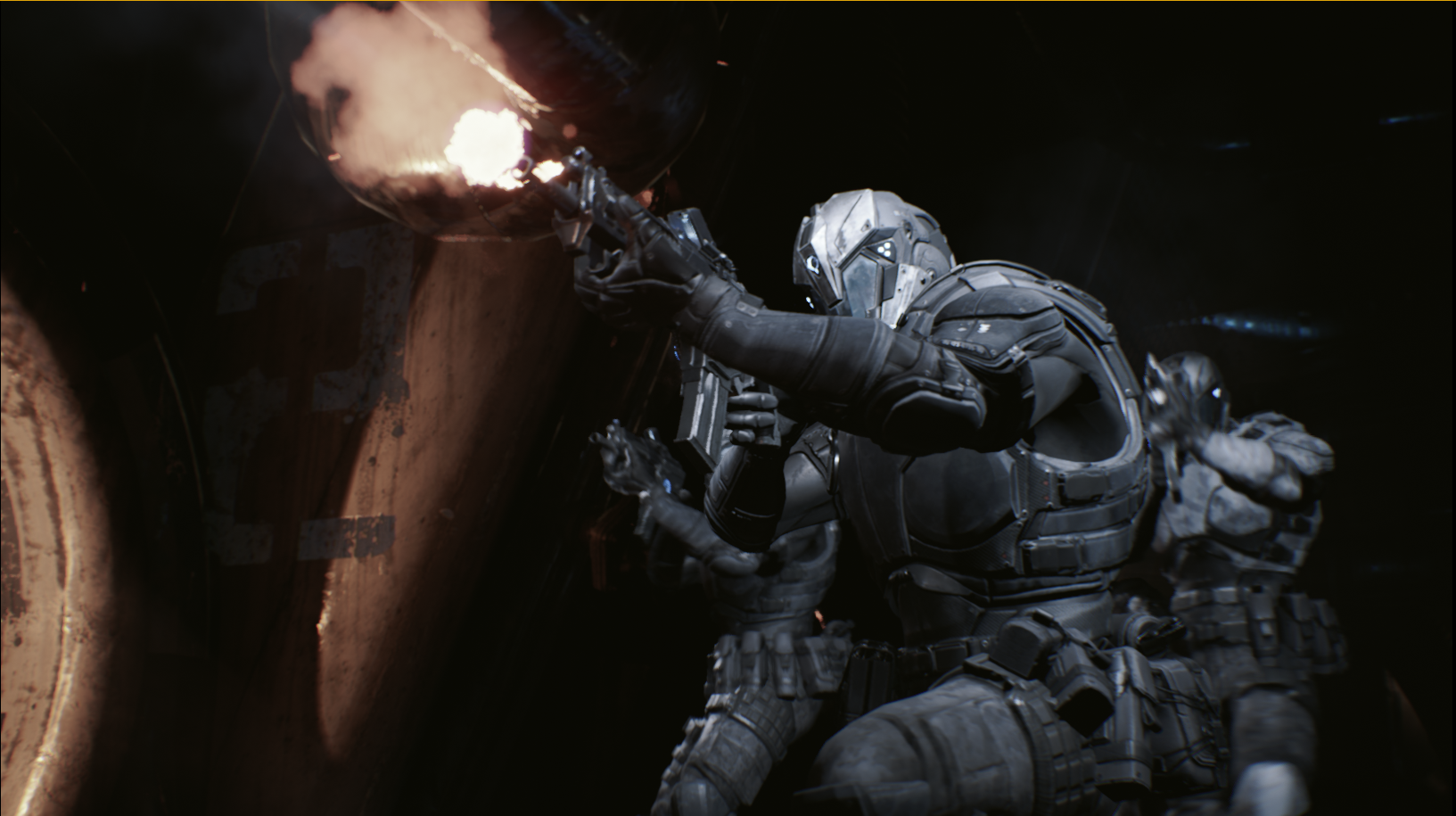}

  \vspace{0.055cm}

  \includegraphics[width=0.48\columnwidth, height=0.27\columnwidth]{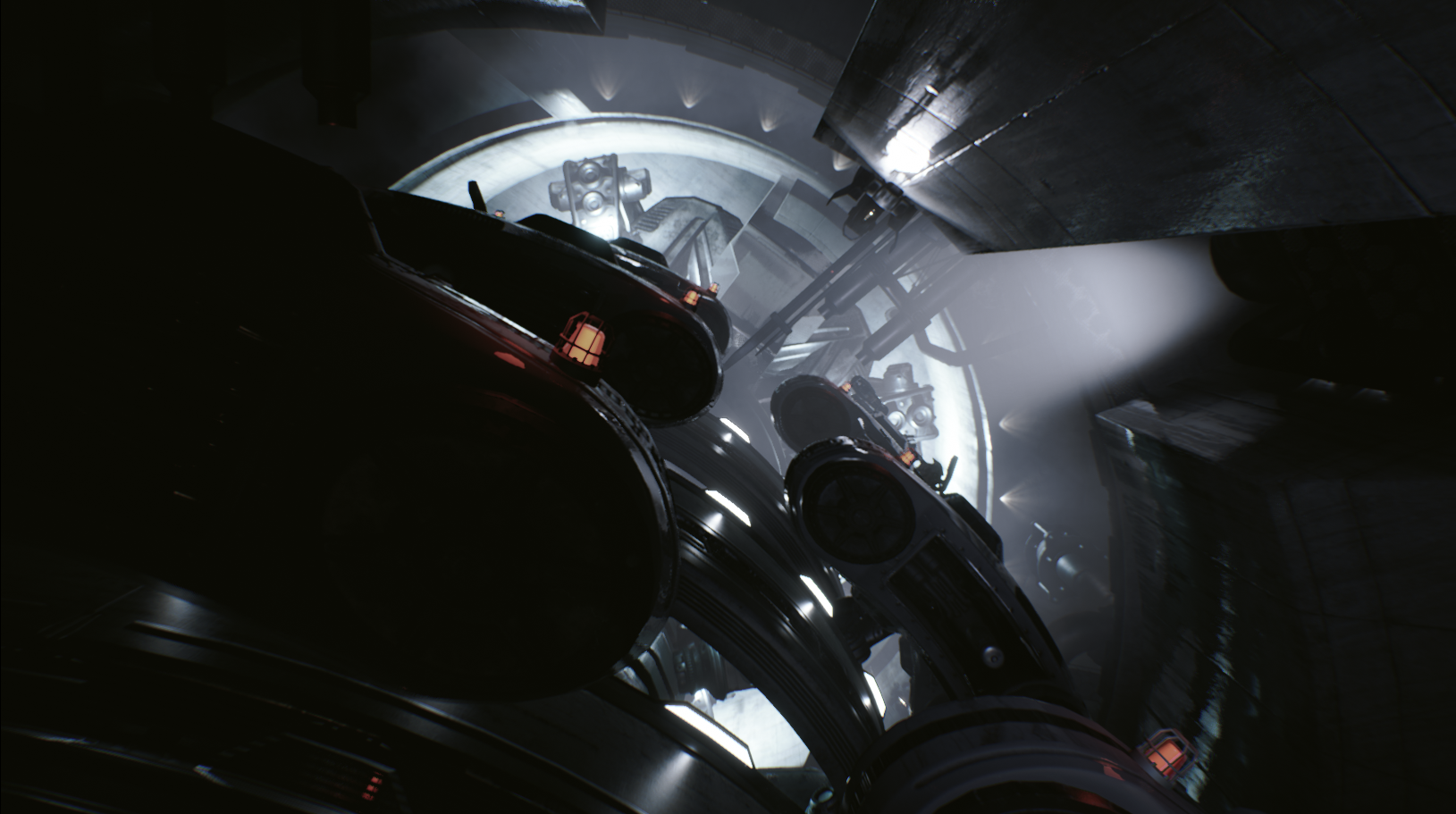}
  \includegraphics[width=0.48\columnwidth, height=0.27\columnwidth]{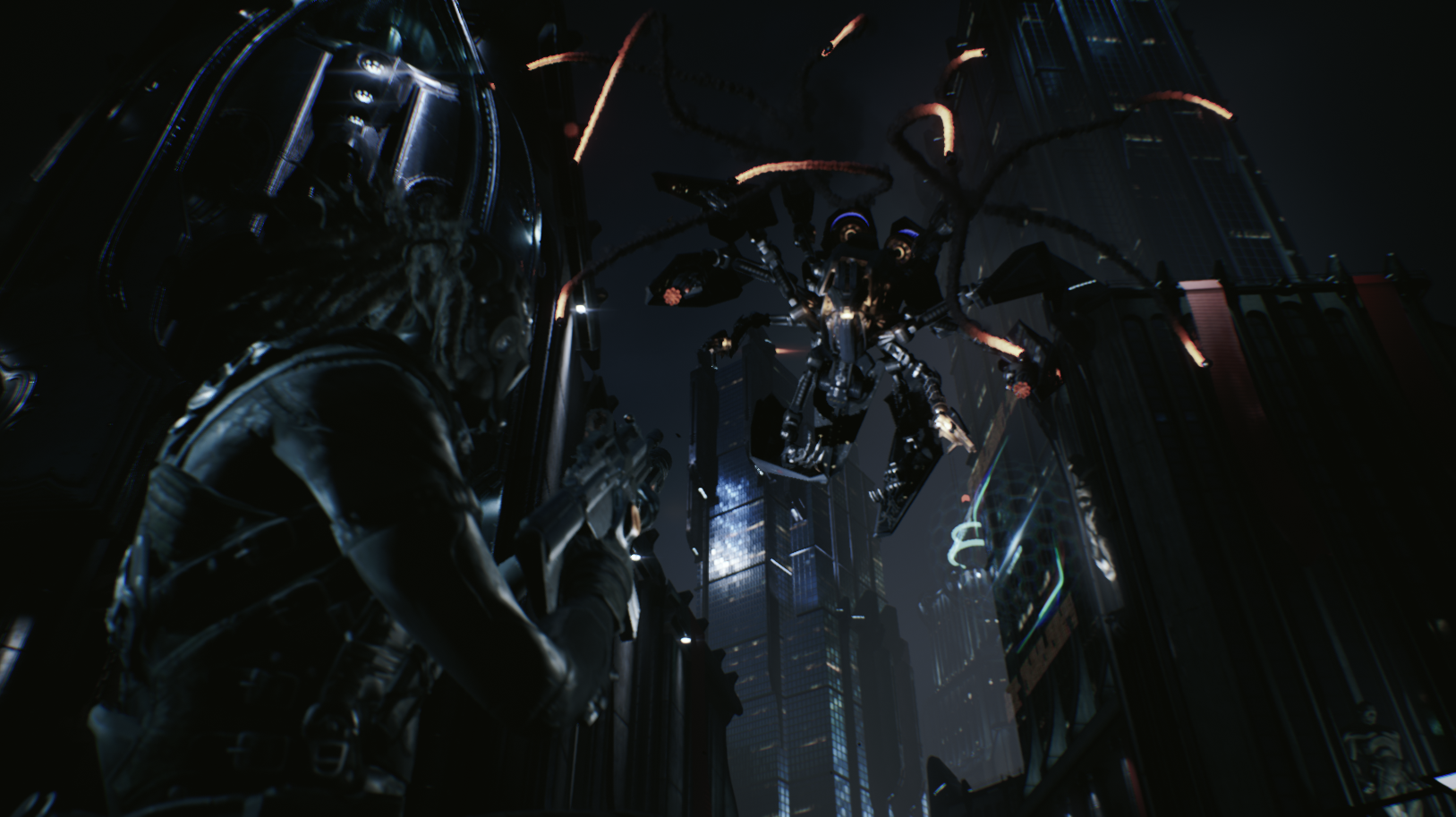}
  \caption{
    Screenshots from the Infiltrator Demo~\cite{EpicGames:2015:ID}.
    We use this demo for our performance evaluation.
    \newtxttwo{Screenshots used with permission of Epic Games Unreal Engine Marketplace.}
  }\label{fig:InfiltratorDemo}
  \Description{A screenshot from the Infiltrator Demo. We use this demo for our performance evaluation.}
\end{figure}

Lastly, we evaluate the impact of our unified design on the runtime performance of GPU code generated by our translator.
We run the Infiltrator Demo\footnote{Epic Games's Infiltrator Demo video: \url{https://youtu.be/dO2rM-l-vdQ}}~\cite{EpicGames:2015:ID} (\figref{InfiltratorDemo}) using both the original UE4 shader code and our rewritten versions and compare the GPU performance in \tabref{Perf_UE4vsUnified}.
These results were produced using a resolution of 2560$\times$1440 on a machine with an Intel Core i7-6700K CPU and an NVIDIA Titan RTX GPU\@. 
As shown in the table, the performance of the shaders ported to our unified environment is comparable to the performance of the original code.

\begin{table}[t]
\centering
\caption[Lines of Code]{GPU performance comparisons for original UE4 shader code vs.\ the versions ported to our unified system.
The table shows the minimum, average, and maximum per-frame execution time in milliseconds for these shaders when running the Infiltrator Demo~\cite{EpicGames:2015:ID}.
These numbers were obtained using benchmarking tools provided by UE4.
}\label{tab:Perf_UE4vsUnified}
\begin{tabular}{lcccccccc}
\toprule
\textbf{Shader}       &\qquad& \multicolumn{3}{c}{\makebox[0pt]{\textbf{Original UE4 Code}}} &\qquad& \multicolumn{3}{c}{\textbf{Our Unified Code}}\\
(runtime in ms)       && Min   & Avg   & Max                                           && Min   & Avg   & Max\\
\midrule
Motion Blur Filter    && 0.06  & 0.18  & 0.70                                          && 0.06  & 0.18  & 0.70\\
Temporal AA           && 0.23  & 0.28  & 0.74                                          && 0.24  & 0.28  & 0.75\\
\bottomrule
\end{tabular}
\end{table}

\section{Limitations and Future Work}\label{sec:Limitations}

Graphics programmers sometimes use specialization parameters to modify struct definitions in HLSL code by using \texttt{\#ifs} to include or exclude certain data member declarations.
They then write corresponding \texttt{\#ifs} throughout the HLSL file whenever they need to access those conditionally defined members.\footnote{This technique is similar to how they conditionally declare uniform parameters based on specialization parameters and, thus, has similar code maintainability downsides.}
While our current system does not support conditional struct definitions, we believe that our idea to co-opt virtual functions for specialization of ShaderClass types can also be applied to specialization of GPU-only struct types.
The key difference is that the data members in a ShaderClass (i.e., uniform and specialization parameters) have the same values for all invocations of a shader program, whereas a GPU-only struct might contain different values per invocation (e.g., if the struct is used as a local variable within a GPU function).
However, as long as all invocations use the same runtime type for the struct (which is equivalent to the HLSL case described above), then the same basic principles can be applied.

In this paper, we have chosen to focus on shaders that align with UE4's \emph{Global Shaders} concept, which are shaders that do not need to interface with the material or mesh systems.
These Global Shaders make up an increasingly large portion of a modern game's shader code and are sufficient to demonstrate the challenges to developing unified shader specialization, as well as how our solutions address these issues.
Therefore, we leave exploration of shaders like UE4's \emph{Material} and \emph{MeshMaterial} shaders as future work, though we believe our method for supporting specialization can extend to these shaders as well.
Similarly, our current implementation only supports compute shaders, but adding support for pixel-type Global Shaders requires only minor changes in the host and GPU backends since pixel- and compute-type Global Shaders are structured comparably.

\begin{sloppypar}
Supporting specialization is one piece of the unified-shader-programming puzzle, and there are many other interesting questions in this space.
For example, MeshMaterial shaders need to coordinate varying parameters between different shader types (e.g., a vertex shader outputs varying parameters that a pixel shader then consumes).
A unified system has the potential to provide a robust mechanism for coordinating this information between different shader types, but the best way to express these cross-shader relationships is an open question (though the \emph{pipeline shader} design~\cite{Proudfoot:2001:ARP} looks promising).
Another area for future work is exploring how a unified environment can better handle data movement and synchronization between host and GPU code.
Our implementation uses UE4's render graph system under the hood for this purpose.
However, a unified system has the opportunity to make better scheduling and memory-transfer decisions because it has a broader view of both the host and GPU code together.
By keeping our system as close to standard \cpp{} as possible, we hope that our ideas can provide a foundation for supporting the crucial shader specialization optimization in unified systems, thereby allowing future works to focus their efforts on other challenges in unified shader programming.
\end{sloppypar}

\section{Conclusion}\label{sec:Conclusion}

In this paper, we have presented the design of a unified shader specialization system in \cpp{}.
By co-opting existing features of the language (attributes and virtual functions) and implementing them with alternate semantics, we are able to provide first-class support for GPU code specialization.
Our system allows programmers to write host and GPU shader code using familiar modularity constructs in \cpp{}, and our source-to-source translator transforms this code into efficient standard \cpp{} and HLSL\@.
\newtxt{Changing the semantics of \cpp{}, even in limited ways, does have some disadvantages.}
\newtxt{Programmers must contend with different semantics in different portions of code, which may increase cognitive load.}
\newtxt{However, we believe the benefits of unified programming and first-class shader specialization far outweigh the downsides of tweaking C++ semantics specifically in GPU shader code.}

\newtxt{Our work demonstrates that unified shader specialization is possible in C++ with only minimally invasive, under-the-hood changes.}
\newtxt{Prior work (\secref{UnifiedShaderProgramming}) has relied on advanced metaprogramming and partial-evaluation features in non-mainstream languages, but these features are absent from the popular languages used in real-time graphics.}
\newtxt{We aspire to bring the benefits demonstrated by these prior works to existing engines today.}
\newtxt{As such, our work is constrained by the use of---and large investment in---\cpp{}.}
\newtxt{Via the co-opting approach, our work helps to bring the benefits of unified shader programming to a widely used language, despite the lack of these advanced language features.}

While our current work focuses on the shader specialization optimization in \cpp{}, we hope the broader lessons can be applied to other programming languages, application domains, and processor types.
Bringing unified shader specialization to other languages may involve co-opting different features, but we think the principles that guided our design are largely transferrable to other, similar languages.
Beyond graphics programming, we believe the strategy of co-opting existing language features can be used to implement the semantics and optimizations needed for other domains and potentially other processor types besides a CPU host and a GPU coprocessor.
This strategy enables programmers to incrementally integrate unified designs while still maintaining compatibility with existing code, which helps to encourage adoption of new ideas and features in existing large-scale systems.

\begin{acks}
We thank Anjul Patney, Chuck Rozhon, Yong He, Brian Karis, Ola Olsson, Andrew Lauritzen, Yuriy O'Donnell, Angelo Pesce, Charlie Birtwistle, Michael Vance, Dave Shreiner, and the anonymous reviewers for guidance, feedback, and technical advice.
Thank you to NVIDIA Corporation for hardware donations and to Intel Corporation for hardware donations and financial support.

\end{acks}

\bibliographystyle{ACM-Reference-Format}
\bibliography{all.bib, tmp.bib}

\end{document}